\documentclass[%
reprint,
 amsmath,amssymb,
 aps,
prl
]{revtex4-1}

\usepackage{graphicx}% Include figure files
\usepackage{dcolumn}% Align table columns on decimal point
\usepackage{bm}% bold math

\usepackage[usenames]{xcolor}

\def\map#1{{{\color{black}#1}}}
\def\cjh#1{{{\color{black}#1}}}

\begin{document}

%\preprint{INT-PUB-18-055}

\title{%\cjh{LIGO search for} compact dark matter objects in the solar system\\
%{\map{LIGO-data search for compact dark matter objects in the solar system}}\\
{\map{Search for compact dark matter objects in the solar system with LIGO data}}}


\author{C. J. Horowitz}
\email{horowit@indiana.edu}
\affiliation{Center for Exploration of Energy and Matter and Department of Physics, Indiana University, Bloomington, IN 47405, USA}

\author{M. A. Papa}
\email{maria.alessandra.papa@aei.mpg.de}
\affiliation{Max Planck Institute for Gravitational Physics (Albert Einstein Institute), D-30167 Hannover, Germany} \affiliation{Department of Physics, University of Wisconsin, Milwaukee, WI 53201, USA}

\author{S. Reddy}
\email{sareddy@uw.edu}
\affiliation{Institute for Nuclear Theory and Department of Physics, University of Washington, Seattle,  WA 98195. }

\date{\today}% It is always \today, today,

\begin{abstract}
Dark matter could be composed of compact dark objects (CDOs).  \cjh{A close binary} of CDOs orbiting \cjh{in the interior of} solar system bodies can be a loud source of gravitational waves (GWs) for the LIGO and VIRGO detectors.  
%An initial search of data from the first Advanced LIGO observing run (O1), sensitive to $h_0\approx 10^{-24}$, rules out close binaries orbiting near the center of the Sun with GW frequencies (twice the orbital frequency) between 50 and 550 Hz and CDO masses above $\approx 10^{-9} M_\odot$. 
\map{We perform the first search ever for this type of signal and rule out close binaries, with separations of order 300 m, orbiting near the center of the Sun with GW frequencies (twice the orbital frequency) between 50 and 550 Hz and CDO masses above $\approx 10^{-9} M_\odot$.} \cjh{ This mass limit is eight orders of magnitude lower than the mass probed in a LIGO search at extra galactic distances.} % well within the range of possible values.}
% \begin{description}
% Secondary publications and information retrieval purposes.
% \item[PACS numbers]
% May be entered using the \verb+\pacs{#1}+ command.
% You may use the \texttt{description} environment to structure your abstract;
% use the optional argument of the \verb+\item+ command to give the category of each item. 
% \end{description}
\end{abstract}

\pacs{Valid PACS appear here}% PACS, the Physics and Astronomy
\maketitle

%\tableofcontents

%\section{\label{sec:intro}Introduction}
Dark matter, or one component of it, could be composed of compact dark objects (CDOs).  These objects are assumed to have small non-gravitational interactions with normal matter and could be primordial black holes, see for example \cite{PhysRevD.95.083508}.  Microlensing observations rule out most of dark matter being made of CDOs with masses between $10^{-7}$ and $15M_\odot$ \cite{Alcock:1995dm,Alcock:2000ph,Tisserand:2006zx,Paczynski:1986}.  In this paper, we focus on CDOs with masses below $10^{-7}M_\odot$ that are not black holes in order to avoid destroying solar system bodies. 

In a previous paper we explored CDOs in the Galaxy \cite{Horowitz_Reddy}, while the LIGO-Virgo collaboration has searched for gravitational waves (GWs) from binaries with masses in the range $0.2-1~M_\odot$ \cite{Abbott:2018oah}. It may be difficult to detect GWs from very low mass CDOs at typical Galactic distances.  Therefore, \cjh{in this letter we search for} GWs from CDOs in the solar system.    Note that there has been considerable discussion of particle dark matter in the Sun, see for example \cite{PhysRevLett.114.081302,PhysRevD.82.083509,PhysRevLett.105.011301}.  Close binaries of CDOs, \cjh{that orbit in the interior of solar system objects}, can be loud sources of GWs for the LIGO/ VIRGO detectors.  At these much shorter distances, far weaker GW signals may be detectable.  Furthermore, these weaker signals only require much smaller radiated powers and therefore the systems may have much longer lifetimes.  This could further increase their detection probability.  {\map{Previous searches for persistent gravitational waves \cite{Abbott:2017pqa,PhysRevD.100.024004,PhysRevD.100.062001,PhysRevLett.123.101101} would have missed these signals because of their unique waveform, as received at the detectors' location.}}

Dark matter in the solar system, and in particular inside Saturn's orbit, is constrained by accurate tracking of spacecraft \cite{Pitjev2013}.  We avoid these limits by focusing on CDOs {\it inside} solar system bodies such as the Sun, Jupiter, or Earth.  In this case the mass of the CDOs would have already been included as a (tiny) part of the observed mass of the body.  In addition, it may be difficult to rule out a number of CDOs at large distances from the Sun where constraints on dark matter in the outer solar system are likely weaker. 

Perhaps occasionally a distant CDO of mass $m_D$ would move into the inner solar system and enter a solar system body.  When this occurs, it is possible that dissipation from unconstrained non-gravitational interactions could trap the CDO in the body and perhaps over a longer time scale bring it towards the center of the body.  Here it may form a close binary with a second trapped CDO.   Furthermore, the lifetime of this binary against GW radiation of a given frequency scales as $m_D^{-5/3}$.  This lifetime can be very long for low mass objects, and ranges from millions of years for $m_D=10^{-8}M_\odot$ to longer than the Hubble time for $m_D=10^{-11}M_\odot$.  We emphasize that a GW source in the solar system, because it is very near, can be detectable even if it has low radiated power.  As a result, the source can live for a very long time, although this also depends on unknown non-gravitational interactions.

We start by estimating the rate of CDO-solar system object collisions assuming dark matter is composed of low mass CDOs.  If the mass density of dark matter near the solar system is $\rho\approx 6\times 10^{-22}$ kg/m$^3$ \cite{0954-3899-41-6-063101}, the number density of CDOs is $n_{D}=\rho/m_D\approx 3\times 10^{-42}$m$^{-3}\times (10^{-10}M_\odot/m_D)$.   Note that $n_D$ increases as $m_D$ decreases.  The rate of CDO collisions with a given body is $\sigma_i\, v_d\, n_{D}$ where $v_d$ is the velocity of the CDO.  For simplicity we consider a single value $v_d\approx 220$ km/s that is adequate for a first rough estimate.  The collision cross section is $\sigma_i=\pi R_i^2[1+(v_{ei}/v_d)^2]$ for a body of radius $R_i$ and escape velocity $v_{ei}$. The total number of collisions during the solar system's lifetime $T_{SS}=5\times 10^9$y is $N_i \approx T_{SS} \sigma_i v_d n_{D}$.  For the Sun the number of collisions is
% \begin{equation}
\cjh{$N_\odot \approx 1.4\times (10^{-10}M_\odot /m_D)$}.
%\end{equation}
If $m_D$ is $10^{-10}M_\odot$ or less, there is a good chance that the Sun has suffered one or more collisions with a CDO during its main sequence lifetime.  For the planets the number of collisions is in general smaller because of their smaller size.  
%For Jupiter $N_J\approx 0.2\times (10^{-12}M_\odot/m_D)$, while for the Earth $N_e\approx  1\times 10^{-3}\times(10^{-12}M_\odot/m_D)$.  
The number of collisions could be higher if the solar system passed through a region of higher dark matter density in the past.  Alternatively, if the Sun somehow has an ``Oort cloud'' of CDOs at large distances, this cloud could be a source of additional collisions.  Finally, CDOs could act as seeds to start the formation of condensed objects and eventually the planets from the solar nebula.  This would naturally explain the presence of CDOs in the planets or other solar system objects.  

What happens to a CDO when it collides with the Sun?  It is possible that there are non-gravitational interactions between the CDO and the conventional matter in the Sun that could help trap the object inside the Sun.  Note that these non-gravitational interactions are poorly constrained.  

{\cjh{Perhaps the weakest point of this scenario is the question of trapping of CDOs inside the Sun or another body. Yet, we believe it is premature to dismiss this possibility. Physical mechanisms that can trap CDOs in the solar system through a combination of non-gravitational interactions and dynamical friction warrant further study. The purpose of the letter is to search for CDOs in the solar system.  These objects could have been trapped, or otherwise formed, in a way that is presently incompletely understood.}} {\map{A positive detection would constitute a landmark result with far reaching consequences. %OR: Clearly, such a detection would mark a watershed moment for all of physics.
}} %We hope that letter serve as a first step towards exploring in detail physical mechanisms needed to trap CDOs in the solar system and alert the broad community of physicists and astrophysicists in the pursuit of discovering dark matter.

If CDOs can be trapped after they collide, and $m_D$ is low enough, it is likely the Sun would now contain one or more CDO.  Two or more CDOs in the Sun could move towards the center (as energy is dissipated) where they may find each other and form a close binary.  Two CDOs orbiting each other at angular frequency $\omega$ will be separated by a distance $r=(2Gm_D/\omega^2)^{1/3}$.  They will radiate GWs at frequency $f_{GW} =\omega/\pi$ that is twice the rotational frequency.  LIGO is insensitive to $f_{GW}$ below about 10Hz.  Therefore two CDOs must come within a distance $[2Gm_D/(10\pi{\rm Hz})^2]^{1/3}$.  This is 300 m for $m_D=10^{-10}M_\odot$.  To avoid touching, each spherical CDO must have an average density greater than $3\pi f_{GW}^2/G$.  This minimum density is $1.4\times 10^{10}$ g/cm$^3$ for $f_{GW}=10$ Hz or $5.6\times 10^{12}$g/cm$^3$ at 200 Hz.  These densities are large but still less than nuclear density $\approx 3\times 10^{14}$ g/cm$^3$.  The density of CDOs is unknown.  Therefore, a LIGO detection would imply high dark matter densities in CDOs, presumably because of appropriate non-gravitational interactions of the dark matter.  

In summary, to generate GWs in LIGO's band a solar system object must contain two or more CDO that are very dense and in a very close binary orbit.  Note that a space based GW detector such as LISA, that is sensitive at lower frequencies \cite{Cornish:2018dyw}, could detect a single CDO moving in the Sun.  This would not require two objects in a close binary orbit and it would not require the CDO to be very dense.  We will discuss this further in a later publication.

%\bigskip

%$<<<<<<<$ SANJAY REDDY adds discussion about dynamical friction and other interactions $ >>>>>>>>>>$

%\bigskip   

Two equal mass CDOs in a circular orbit will produce a GW strain $h_{jk}$ of,
\begin{equation}
    h_{jk}=\frac{2G}{c^4d}\frac{d^2}{dt^2}I_{jk}^{TT}\, .
\end{equation}
Here $d$ is the distance to the source and $I_{jk}^{TT}$ is the transverse traceless quadrupole moment \cite{Poisson_Will}. The intrinsic strain amplitude is \cjh{$h_0 \propto h_{xx}-h_{yy} = 4G m_Dr^2\omega^2/(c^4 d)$}.
%    \label{Eq.h0}
%\end{equation}
%where $r$ is the distance between the CDOs. % The angular frequency $\omega$ is related to the gravitational wave frequency $f_{GW}=\omega/\pi$.  
Using Kepler's law $r^3=2Gm_D/\omega^2$ to eliminate $r$ yields,
\begin{equation}
h_0=\frac{2^{8/3}\pi^{2/3}}{c^4d}(Gm_D)^{5/3}f_{GW}^{2/3}\,.
\end{equation}
For objects inside the Sun with $d=1.5\times 10^{11}$ m, we have,
\begin{equation}
h_0=1.4\times 10^{-9} \Bigl(\frac{m_D}{M_\odot}\Bigr)^{5/3}\Bigl(\frac{f_{GW}}{200 {\rm Hz}}\Bigr)^{2/3}.
\label{Eq.h0sun}
\end{equation}
The strain amplitude $h_0$ in Eq.~\ref{Eq.h0sun} is dramatically larger than many galactic GW sources because the distance $d$ is very small, only $10^{-8}$ kpc.  \cjh{Note that the impact on the GW signal of possible non-gravitational interactions (of the CDOs with conventional matter) remains to be explored.}
%For objects orbiting inside the moon, assuming $m_D\ll$ the mass of the moon, we have,
%\begin{equation}
%h_0=5.5\times 10^{-7} \Bigl(\frac{m_D}{M_\odot}\Bigr)^{5/3}\Bigl(\frac{f_{GW}}{200 {\rm Hz}}\Bigr)^{2/3}.
%\label{Eq.h0moon}
%\end{equation}
%Alternatively, for low mass objects orbiting near the center of the Earth we have,
%\begin{equation}
%h_0=3.3\times 10^{-5} \Bigl(\frac{m_D}{M_\odot}\Bigr)^{5/3}\Bigl(\frac{f_{GW}}{200 {\rm Hz}}\Bigr)^{2/3}.
%\label{Eq.h0earth}
%\end{equation}
%Again, this signal could be large, even for very low $m_D$, because the distance is extremely small.

%\bigskip

%$<<<<<<<<<<<<<<$

%\smallskip
%\section {Simple search}
%Maria Alessandra please add discussion of search. Take the space you need.  We can cut elsewhere.  You could include a figure showing the detector noise data for some spectral range and perhaps a blow up of one line that nearly passes the search criteria...  From the search we conclude mass limits as shown in figure  \ref{Fig1}.  [Please tell me how to revise this figure.]
Since the signal amplitude is potentially so large we perform a simple search in the frequency band 50-550 Hz on the most sensitive publicly available gravitational wave data, at the end of 2018: that from the first Advanced LIGO observing run (O1) \cite{O1data,thankslosc}. 
We assume that the center of mass of the binary is nearly at rest with respect to the Sun. If the center of mass of the binary were also at rest with respect to the gravitational wave detectors, the gravitational wave signal at the detectors would be a pure tone at the frequency $f_{GW}$.  However, because of the relative motion between the CDOs and the detectors due to the rotation of the Earth, the signal at each detector is Doppler-modulated in a way that varies in time depending on the relative velocity between the CDOs center of mass and the detector. The modulation has a periodicity of a day and the maximum Doppler shift for a signal at a frequency $f_{GW}$ is $10^{-6} f_{GW}$. 

The frequency spacing of a Fourier transform of data covering a period of time $T_{\textrm{FFT}}$, is $\delta f = \frac{1}{T_{\textrm{FFT}}}$. Any signal whose instantaneous frequency does not vary by more than $\delta f$ during $T_{\textrm{FFT}}$ will appear like a monochromatic signal, i.e. with the signal power concentrated at the frequency bin(s) corresponding to the signal frequency \cite{allenpapa2002}. We will refer to this as the ``signal peak".

The longer $T_{\textrm{FFT}}$ is, the higher is the signal peak. As the $T_{\textrm{FFT}}$ increases and the $\delta f$ decreases, because the signal frequency varies, there will come a point when the signal's instantaneous frequency during $T_{\textrm{FFT}}$ is not confined to one or two bins, but moves over more bins. As a result the signal power is not concentrated in a single bin but spread over more and hence decreases in amplitude in all of them compared to what it would be if the frequency did not vary. For frequency variations due to the Doppler shift of the Earth's spin, the maximum $T_{\textrm{FFT}}$ such that the instantaneous frequency does not shift by more than half a frequency bin is \cite{Krishnan:2004sv,Frasca_2005}
%:SFT
\begin{equation}
T_{\textrm{FFT}}^{\textrm{max}} = \frac{4.7 \times 10^{4}}{\sqrt{f_{GW}}} {\textrm{s}}.
\label{eq:DopplerBin}
\end{equation}
For $f_{GW}=550$ Hz this yields $T_{\textrm{FFT}}^{\textrm{max}}=2.0\times 10^3$ s; hence we take Fourier transforms of the LIGO O1 data over time intervals that are 1800 s long. This will give maximum sensitivity to a signal during each half hour, without having to perform any demodulation.

The instantaneous frequency of the signal does not move by more than half a  frequency bin during each half hour, but it does in general move over the course of the observation time. With a frequency resolution of $\delta f ={1/1800}$ Hz, the signal frequency $f_0$ such that $f_0 \times 10^{-6} = \delta f$ is about 555 Hz. Signals with frequencies higher than this, would appear in different bins depending on the observation time. An optimal detection procedure would require tracking of such peaks and would be more involved than what we do here.  Therefore 550 Hz is the highest frequency we consider.

The FFTs are prepared following the standard procedure used for the Einstein@Home searches (see for instance \cite{Abbott:2017pqa}), including a procedure that eliminates very loud time-domain disturbances from the data \cite{gating}. For this reason we use the standard name for these FFTs computed over short time intervals, i.e. SFTs (Short timebaseline Fourier Transforms)\cite{SFTspecs}. 

In total we produce 3507 SFTs from the Hanford detector data (LHO) and 2889 from the Livingston detector (LLO). We indicate the SFT data with $\tilde{x}_{\alpha,k}^I$, where $\alpha$ is the SFT order-number, $k$ is the frequency index and $I=H$ or $L$ indicates the detector. The conventions for the SFT data are given in \cite{SFTspecs}. 

\begin{figure}[ht]
\smallskip
\includegraphics[width=1.1\columnwidth]{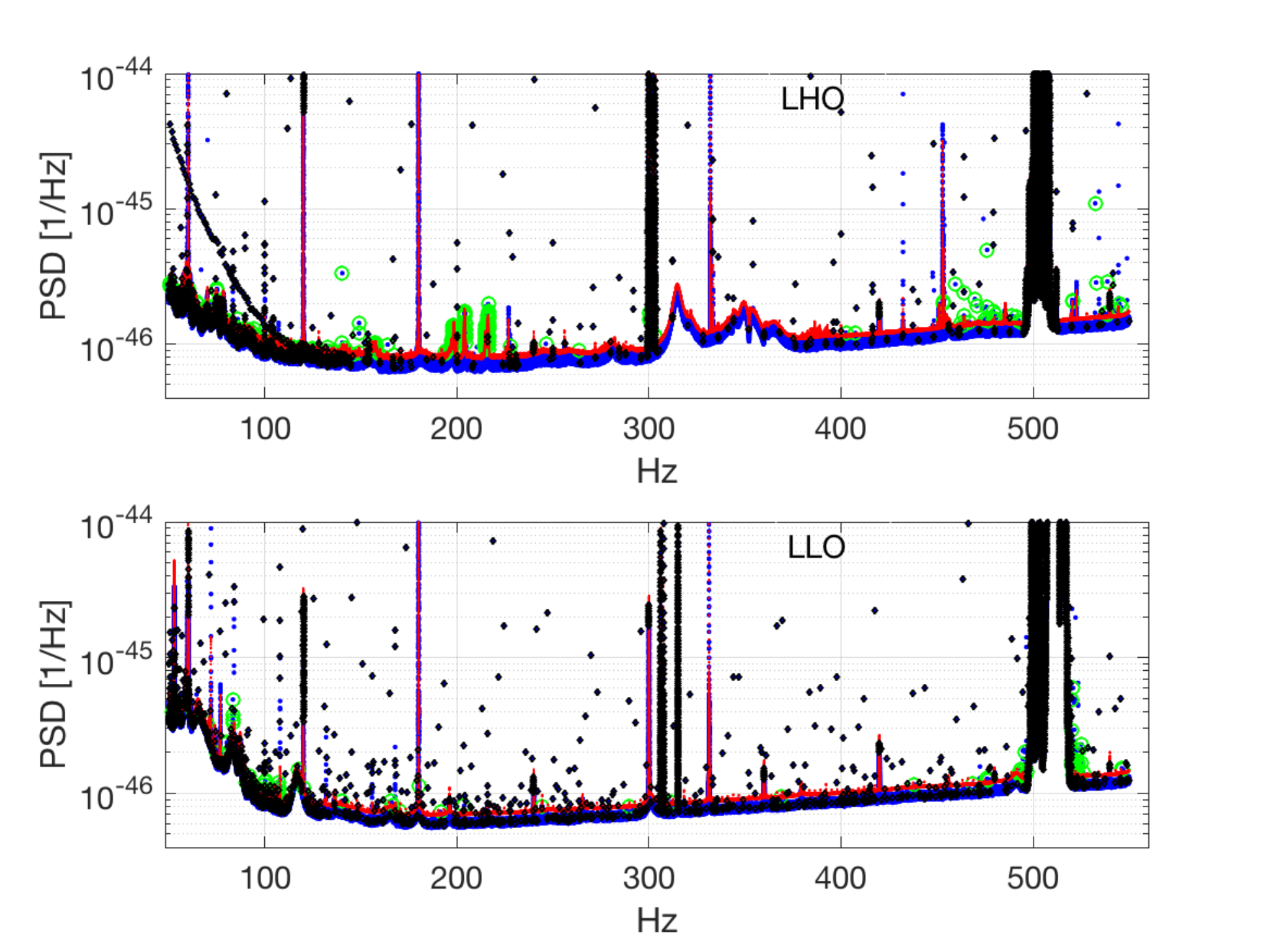}
 \caption{(Color online) 
Power spectral density (PSD) for the Hanford (LHO) and Livingston (LLO) detector, estimated with the data used for this search. The blue points are the original data set. The black points indicate frequencies that are vetoed based on known line lists \cite{Covas:2018oik}. The red line is the threshold used to select outliers. The green circles indicate the outliers which fall at unvetoed frequencies.}
\label{fig:PSDs}
\end{figure}

For each detector I and each SFT $\alpha$ we compute the power spectral density as a function of frequency $k$ 
\begin{equation}
\label{eq:ASD}
p_{{\alpha,k}}^I = {\frac{2}{T_{\textrm{SFT}}} |\tilde{x}_{\alpha,k}^I |^2},
\end{equation}
with $T_{\textrm{SFT}}=1800$ s, and its average over SFTs:
\begin{equation}
\label{eq:AvASD}
p_{{k}}^I = \frac{1}{N^I}\sum_\alpha p_{{\alpha,k}}^I,
\end{equation}
with $N^I$ being 3507 and 2889 for the Hanford and Livingston detector respectively. Figure~\ref{fig:PSDs} shows $p_{k}^I$ for the two LIGO detectors. 

\begin{figure}[ht]
\smallskip
\includegraphics[width=1.1\columnwidth]{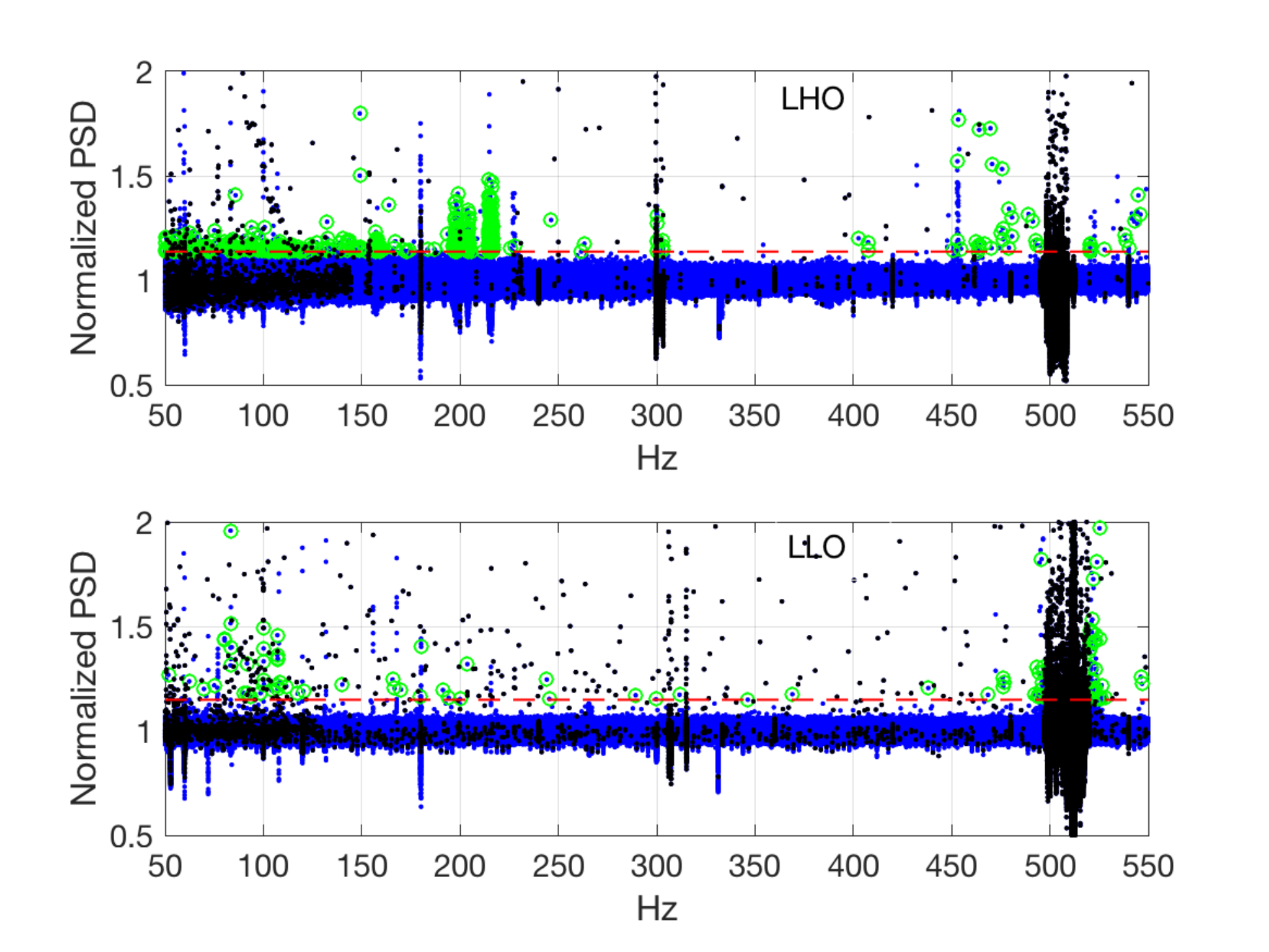}
 \caption{(Color online) 
Normalised average power spectral densities. The blue points are the original data set. The black points indicate frequencies that are vetoed based on known line lists \cite{Covas:2018oik}. The magenta dashed line is the 8 $\sigma$ threshold. The green circles indicate the points above threshold that come from frequency bands whose statistic is within the accepted range (the mean is within 2.5$ \sigma$s of the Gaussian noise value.).}
\label{fig:NormalisedPSDs}
\end{figure}

A loud CDO signal will appear as a peak in the power spectral density of both detectors at the same frequency. %Due to the fact that the signal is not perfectly monochromatic the signal peak might be in two different bins over the course of the observation producing in the average power spectral density a loss of at most $\sim$ 50\% in the peak power.
Since the noise level in the detectors varies with frequency, we do not set a fixed threshold but rather we identify signal peak candidates as outlier values of the {\it normalized}  average power spectral density. We normalize the average power spectral density in each detector with the average of the  running median power spectral densities, with a window of 21 bins. This detection statistic has an expected value of 1.0 and a standard deviation of $\simeq$ 0.024 for Gaussian stationary noise data. Since the detector noise is neither Gaussian not stationary, we do not expect that the normalized average power spectral density values (shown in Fig.~\ref{fig:NormalisedPSDs}) generally follow the predicted Gaussian-noise behaviour.  This makes it hard to assess what an outlier is.

%We hence expect that a monochromatic signal with intrinsic amplitude at or above $10^{-23}$ would be clearly visible as a peak in the detectors' amplitude spectral density at or above $5\times 10^{-23}~{1\over{\sqrt{\textrm{Hz}}}}$. 

\begin{figure}[ht]
\smallskip
\includegraphics[width=.9\columnwidth]{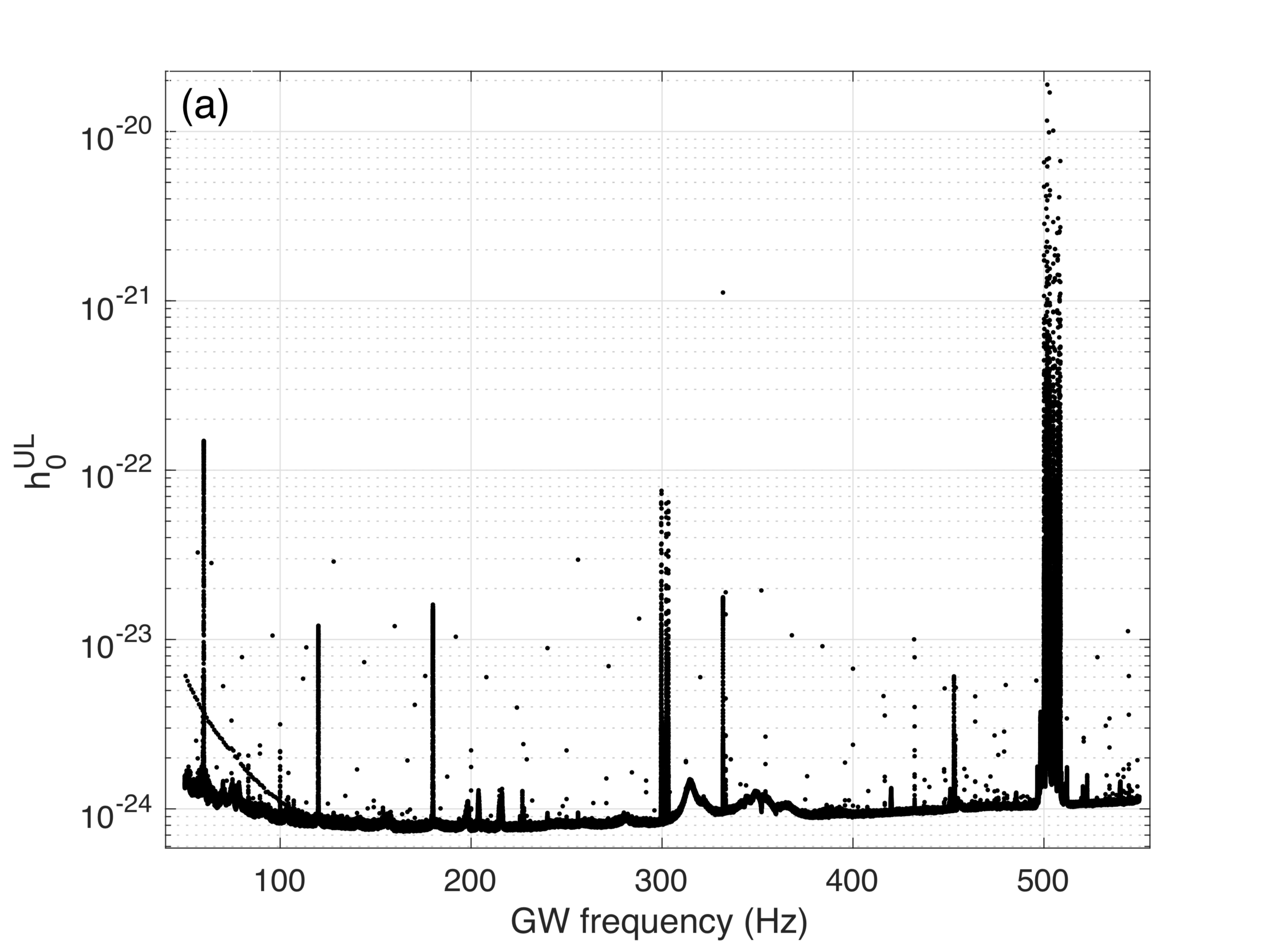}
\includegraphics[width=.9\columnwidth]{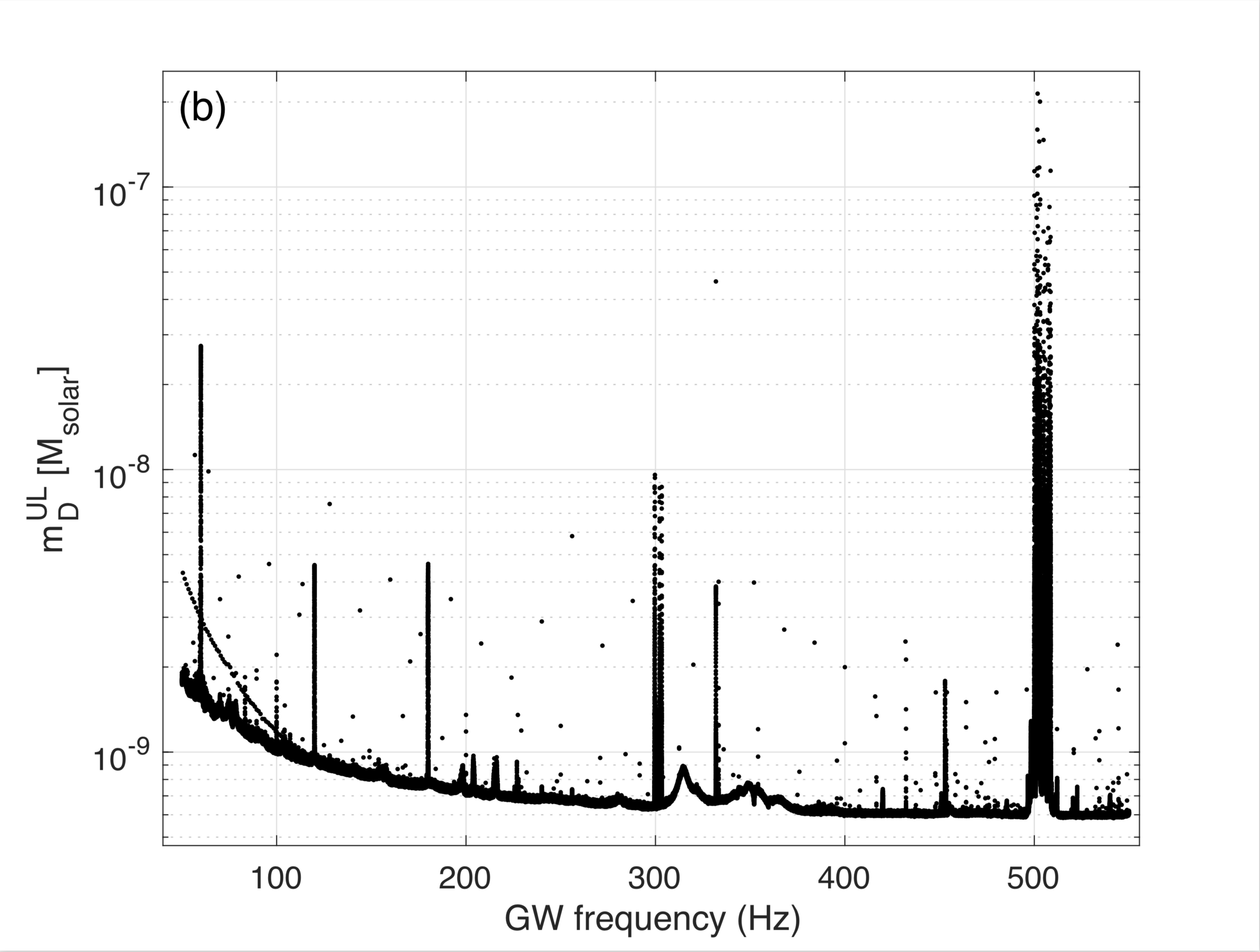}
 \caption{Intrinsic gravitational amplitude upper limits (a) and upper limits on the mass of compact dark matter object (CDO) binaries orbiting at the center of the Sun (b), as a function of the gravitational wave signal frequency. Based on this search, we can exclude values above these curves.
}
\label{fig:ULs}
\end{figure}

As shown in Figs.~\ref{fig:PSDs} and \ref{fig:NormalisedPSDs}, the power spectral density of the O1 data displays a number of lines. This is well known and was studied in depth by the LIGO team \cite{Covas:2018oik}. As a result a large part of these lines have been identified as due to local disturbances acting on the instruments. We exclude from our analysis any frequency bin that is in the range identified by \cite{Covas:2018oik} and references therein to LOSC pages. On this data we further only consider sets of 90 bins whose average of the normalized power spectral density (after removing the highest value) is within 2.5 standard deviations of the Gaussian-noise expectations. These cuts remove between 12\% (LLO) and 18\% (LHO) of the total number of bins. 
%StatsInBlocks on 90 bins
For the remaining frequencies we set a threshold corresponding to 8 Gaussian standard deviations and cluster together bins that are not more distant than 3 frequency bins. We find 83 such clusters in the LLO data and over 466 in the LHO data, reflecting the different noise conditions of the two detectors. We require that a signal be present in both detectors at similar frequencies and look for clusters in the LIGO and Hanford data that overlap in frequency. We find none and conclude that there is no clear detection of this kind of signal.  We proceed to set upper limits on the intrinsic amplitude of such gravitational wave signals $h_0^{UL}$ (Fig.~\ref {fig:ULs}a), and then translate these into limits on the CDOs mass (Fig.~\ref {fig:ULs}b).

%We consider only the frequency bins that are free of such known disturbances and in those bins we determine the average amplitude spectral density over the detectors, $p_k$. 

A loud monochromatic signal of amplitude $h_0$  optimally oriented with respect to the detector will produce a peak in the power spectral density of about $p_0=h_0^2 {T_{\textrm{SFT}}}$. The frequency of the peak will be at the frequency of the signal. If the detector were not optimally oriented with respect to the signal, as is generally the case, the signal amplitude spectral peak will be lower than $p_0$ by a factor of a few, and this factor will be slightly different in each detector. For a signal coming from the Sun, on average a reduction by a factor of $\simeq 7.8$ is expected with respect to optimal orientation.
%, with the loss being about 15\% higher for LHO than for LLO, because LHO is at located on Earth at a higher latitude. 
% <d_2^2>_{alpha,psi,iota}=4/5 e2(delta) h_0^2 T/Sh
% <e2(delta)>= 0.16 for LHO and 0.2 for LLO
% 4/5=0.8  => 0.16 x 0.8 = 0.128 => 1.0/0.128 = 7.8
% 0.16 x 0.8 x 0.5 = 0.064 its inverse = 15.625 and the sqrt of the inverse 3.9528
%sqrt(1800)/3.9528 = 10.73

We translate the observed power spectral density values into intrinsic gravitational wave amplitude upper limits $h_0^{UL}$ at different frequencies as follows: 
\begin{equation}
  h_0^{UL}(f_k)=\sqrt { \frac{15.6 ~ {{{\textrm{max}_I}}[p^I_k] }}  {T_{SFT}}}=
  \frac{1}{10.7} \sqrt{{{\textrm{max}_I}}[p^I_k]} 
  %\left[ {1800\over{T_{SFT}}}\right]^{1/2}
  ,   \label{eq:h0UL}
\end{equation}
where the factor of 15.6 includes an additional reduction in signal power for higher confidence for the least favourable polarizations and inclinations of the source. These upper limits are plotted as a function of the gravitational wave frequency in Fig.~\ref{fig:ULs}a. We translate them into upper limits on the mass of CDOs, obtaining the curve shown in Fig.~\ref {fig:ULs}b.  For many frequencies the limit is less than $10^{-9}M_\odot$ and the most stringent upper limit on the CDO mass is $5.8\times 10^{-10}~M_\odot$ at $\simeq 525.5$ Hz.  \cjh{These mass limits are eight orders of magnitude lower than a previous LIGO search \cite{Abbott:2018oah}.}

We have excluded from the search
%,iu8i 
frequency bins that appear polluted by spurious noise that we could not model. We did this, because we would not have been  able to assess the significance of a signal candidate from such frequencies. We are however able to place upper limits on the signal amplitude, even at these frequencies. At frequencies with large spectral lines the upper limits are not as constraining as they are in ``quieter'' bands, but they are still valid.

This is a very simple search that has reached a modest sensitivity depth \cite{Behnke:2014tma} of $\sim 10$ Hz$^{-1/2}$. Other searches for continuous signals from known objects typically reach sensitivity depths of \cite{PhysRevD.98.084058} a few hundred Hz$^{-1/2}$ and broad parameter space searches reach a depth of several tens Hz$^{-1/2}$. \cjh{Reaching these sensitivity depths is possible and this paper paves the way for significantly more sophisticated searches that could probe a broad ($\approx 2$ kHz) range of frequencies and CDO masses below $10^{-10}M_\odot$.}

%{\map{Reaching these sensitivity depths is possible, but requires significantly more sophisticated searches: with coherent time-baselines much longer than 1800s, coherently combining the data from both detectors and appropriately summing the results from the different searches. Such searches could probe a broad ($\approx 2$ kHz) range of GW frequencies and CDO masses below $10^{-10}M_\odot$. This paper paves the way to these future investigations.}}

%%%%%%%%%%%%%%

%\bigskip

%\smallskip
%Also we could discuss eccentric orbits and orbits of the CDO binary center of mass about the center of the solar system object.  These orbits have low frequencies given by the central densities of the solar system bodies.  I can add this (CJH).
 
%\smallskip
%$>>>>>>>>>>>>>>$

%\bigskip

In addition to the Sun, \cjh{we plan to} search for GWs from CDOs in the Earth and Jupiter.  Jupiter's gravity influences much of the solar system and many objects, such as Comet Shoemaker Levy 9, have collided with it.  %On average, Jupiter is about 5 astronomical units from Earth so our mass limit for CDOs in close binary orbits in Jupiter is a factor of about $5^{3/5}=2.63$ larger than the limit for the Sun that is shown in figure ~\ref{Fig1}. 
%We plan to search for GWs from CDOs in Jupiter.  
%If a GW signal is observed, its origin in Jupiter can be confirmed by verifying that the amplitude varies with one over the Earth to Jupiter distance.  
%We also plan to search for GWs from CDOs near the center of the Earth.  
\cjh{For the Earth, the very small distance to the source (that may be less than a GW wavelength) likely will allow sensitivity to the lowest CDO masses.}  

In conclusion, dark matter could be composed of compact dark objects (CDOs).  A close binary of CDOs orbiting {\it inside} solar system bodies can be a loud source of gravitational waves (GWs).  We have performed an initial search for GWs from the Sun, using data from the first Advanced LIGO observing run (O1), that reached a sensitivity of $h_0\approx 10^{-24}$.  This search rules out close binaries of CDOs orbiting near the center of the Sun with GW frequencies (twice the orbital frequency) between 50 and 550 Hz and CDO masses above $\approx 10^{-9} M_\odot$. 

\acknowledgements{CJH thanks the Max Planck Institute for Gravitational Physics in Hannover for its hospitality. CJH is supported in part by DOE grants DE-FG02-87ER40365 and DE-SC0018083. SR acknowledges support from the US Department of Energy Grant No. DE-FG02-00ER41132.  This research has made use of data, software and/or web tools obtained from the Gravitational Wave Open Science Center (https://www.gw-openscience.org), a service of LIGO Laboratory, the LIGO Scientific Collaboration and the Virgo Collaboration. LIGO is funded by the U.S. National Science Foundation. Virgo is funded by the French Centre National de Recherche Scientifique (CNRS), the Italian Istituto Nazionale della Fisica Nucleare (INFN) and the Dutch Nikhef, with contributions by Polish and Hungarian institutes.}
\bibliographystyle{apsrev}
%\bibliography{DarkRattles}

\end{document}